\documentclass[11pt]{article}
\usepackage{amsmath}
\usepackage{amssymb}
\usepackage{bm}
\usepackage[dvips]{graphicx}
\usepackage{color}

\begin{document}

\begin{titlepage}

 \setcounter{page}{0}

 \begin{flushright}
  KEK-TH-1174 \\
  OIQP-07-08 \\
  YITP-07-51
 \end{flushright}

 \vskip 5mm
 \begin{center}
  {\Large\bf Higher-spin Gauge and Trace Anomalies \\in Two-dimensional
  Backgrounds}  

  \vskip 15mm

 {\large Satoshi Iso$^{1,}$\footnote{\tt satoshi.iso@kek.jp}, 
  Takeshi Morita$^{2,}$\footnote{{\tt mtakeshi@yukawa.kyoto-u.ac.jp,
  takeshi@theory.tifr.res.in}, \\
  Present address: Department of Theoretical Physics, Tata Institute of
  Fundamental Research, Homi Bhabha Road, Mumbai 400005, India. 
  } 
  and Hiroshi Umetsu$^3,$\footnote{\tt hiroshi\_umetsu@pref.okayama.jp} }

  \vspace{5mm}
  $^1${\it Institute of Particle and Nuclear Studies \\
  High Energy Accelerator Research Organization (KEK) \\
  Oho 1-1, Tsukuba, Ibaraki 305-0801, Japan 
  }
 
  \vspace{5mm}

  $^2${\it Yukawa Institute for Theoretical Physics, Kyoto University
  \\
  Kyoto 606-8502, Japan}
  
  \vspace{5mm}
  $^3${\it Okayama Institute for Quantum Physics \\
  Kyoyama 1-9-1, Okayama 700-0015, Japan} 
 \end{center}

\vskip 30mm 

 \centerline{{\bf{Abstract}}} 

 \vskip 3mm 

Two-dimensional quantum fields in electric and gravitational backgrounds can
be described by conformal field theories, and hence all the physical
(covariant) quantities can be written in terms of the corresponding
holomorphic quantities.  In this paper, we first derive relations between
covariant and holomorphic forms of higher-spin currents in these
backgrounds, and then, by using these relations, obtain higher-spin
generalizations of the trace and gauge (or gravitational) anomalies up to
spin 4.  These results are applied to derive higher-moments of Hawking
fluxes in black holes in a separate paper \cite{IMU5}.
\end{titlepage}

\newpage

\section{Introduction}
\setlength{\baselineskip}{7mm}

\setcounter{footnote}{0}
\setcounter{equation}{0}
Hawking radiation is a universal quantum effect which arises in the
background spacetime with event horizons \cite{Hawking1,Hawking2}. 
Such universal behavior arises because  fields in 
black hole backgrounds can be reduced to an infinite
set of two-dimensional conformal fields near the horizon.
The emergence of conformal symmetries near the horizon was first
emphasized in \cite{conformal} and used 
to derive the Hawking radiation based on gauge or 
gravitational anomalies \cite{Robinson,IUW1}.
The anomaly method has been applied to rotating black holes
\cite{IUW2,Soda} and various others.
Such conformal structure near the horizon is also used to derive the 
higher-spin (HS) currents of Hawking radiation \cite{IMU2,IMU3}
by examining conformal transformation properties of
these HS currents.

The above derivation of the HS fluxes
is based on  the fact that two-dimensional
quantum fields can be described by conformal field theories
even in the presence of the electric and gravitational
backgrounds, and hence all the physical 
quantities are written in terms of the conformal,
i.e. holomorphic and anti-holomorphic, quantities.
The HS currents used in  \cite{IMU2,IMU3}
are the holomorphic currents.
They are holomorphic functions and 
 different from the $(u \cdots u)$-component  of  the ordinary 
covariant currents by some functions of 
the electric and gravitational backgrounds.
These differences are responsible for the 
conformal transformation properties of the 
conformal currents.
In the simplest case of the energy-momentum (EM) tensor
in the gravitational background,
it is well-known
 that we can define 
the holomorphic EM tensor $t(u)$ from the 
original covariant EM tensor  $T_{uu}$ by
\begin{equation}
t(u) = T_{uu} -\frac{c}{24 \pi} \left( 
 \partial_u^2 \varphi - \frac{1}{2}(\partial_u \varphi)^2\right), 
\end{equation}
where $c$ is the central charge and $\varphi$ is the conformal factor 
of the gravitational background. This relation gives 
the transformation property of $t(u)$ under conformal transformations.

In this paper, we generalize this relation to all the HS
currents in electric and gravitational backgrounds.
This gives a further justification of our analysis in 
\cite{IMU2} and \cite{IMU3}. 

In section 2, we first review the relation between covariant 
and holomorphic forms of the $U(1)$ current and the 
EM tensor in electric and gravitational backgrounds.
For the case of EM tensor, this relation 
can be obtained from the conservation
equations for EM tensor $\nabla_\mu T^\mu_\nu = F_{\mu \nu} J^\mu$ 
and the trace anomaly $T^\mu_\mu =cR/24\pi$.
Note that $T_{\mu\nu}$ denotes the matter EM tensor and
therefore it is not conserved by itself in the electric 
background. 

In section 3, we generalize them to HS currents.
Here we construct higher-spin ($W_{1+\infty}$) currents from
two-dimensional fermion fields in the electric and
gravitational backgrounds.
Since we do not know either
conservation equations or trace anomalies for HS
currents at the beginning, we cannot start from these equations.
Instead we will take the following procedure to obtain the
relations between covariant and holomorphic HS currents.
The original fermion field $\psi$ transforms covariantly
under gauge and local Lorentz transformations.
We will construct  covariant HS currents 
by regularizing the fermion  bilinears
$\partial^n \psi^\dagger(x)  \partial^m \psi(x) $
in the covariant way under gauge and general coordinate transformations.
On the other hand, we can define a new fermion 
field $\Psi$ which is holomorphic in the electric
and gravitational backgrounds, and by using it, 
we construct a holomorphic form of the HS currents.
After defining these two types of HS currents,
we give  relations between the covariant and holomorphic
forms of HS currents.

In section 4, by
using the relations between covariant and conformal HS currents in
section 3, 
we obtain  conservation equations 
and trace anomalies for the HS currents.
This is the inverse step compared
to the derivations of the holomorphic $U(1)$ current and EM tensor
in section 2.
We show that the relations in section 3 and
some assumptions for the currents are sufficient to determine the
explicit forms of trace anomalies for HS currents.
For the classically traceless spin 3 current $J^{(3)}_{\mu \nu \lambda}$, 
it acquires the following quantum correction: 
\begin{align}  
{J^{(3)\mu}}_{\mu \nu} = \frac{\hbar}{12\pi} \nabla_\mu {F^\mu}_{\nu}.
\end{align} 
This is considered as a spin-3 generalization of the trace anomaly for
EM tensor. For the spin 4 current $J^{(4) }_{\mu \nu\rho \sigma}$, it is
classically traceless but it acquires the quantum anomaly given by
 \begin{align} 
 {J^{(4)\mu }}_{\mu \nu\rho}
  =-\frac{\hbar}{160\pi} \nabla_\nu \nabla_\rho R
  +\hbar g_{\nu\rho}\left[
  \frac{1}{160\pi}\nabla^2 R+\frac{1}{24\pi}\left({\tilde{F}}^2
  -\frac{13}{120}R^2  \right)  \right] .
 \end{align} 
A generalization to higher spins than 4 is also possible
but the calculation becomes more complicated.

In section 5, we consider a chiral theory where
the central charges in the left and right handed sectors
are different. In this case, 
we can obtain a generalization of the 
gauge(or gravitational) anomalies for higher-spin currents.  
We first review how we get the
gravitational anomaly from the relations obtained  in section 3,
and then generalize it to spin 3 and 4 currents.
For the spin 3 current, the generalization of the 
gauge anomaly becomes
\begin{align}
\nabla_\mu J^{(3) \mu}_{\nu \rho} = \cdots \pm
\frac{\hbar}{96\pi} 
\left( \epsilon_{\nu \sigma} \nabla^\sigma \nabla_\mu F^\mu_\rho
+ \epsilon_{\rho \sigma} \nabla^\sigma \nabla_\mu F^\mu_\nu
-g_{\nu \rho}\epsilon_{\alpha \sigma} \nabla^\sigma \nabla_\mu F^{\mu\alpha}
\right).
\end{align}
Here $\cdots$ represents classical violation of the conservation equation for
matter currents in the electric and gravitational background. $+(-)$
corresponds to the right (left) handed fermion.

These results can be applied to derive the HS fluxes 
of Hawking radiation.
The relations between covariant and conformal HS currents 
obtained in section 3 
 provide another 
derivation of fluxes of HS currents in Hawking radiation. 
These relations are equivalent to  solving
the conservation equations and trace anomaly equations
for HS currents. Hence the derivation gives a 
generalization of the Christensen and Fulling's method \cite{CF},
in which the conservation equation of the EM tensor 
and the trace anomaly equation are solved
with the regularity condition at the horizon.
On the other hand, 
as we see in section 5, these relations 
can be rewritten as a generalization of the 
gauge anomaly. 
By applying these anomaly equations to black holes,
it gives a generalization of the anomaly method \cite{Robinson, IUW1}
(see also appendix of \cite{IMU1} and \cite{Banerjee}).
These two derivations also clarify some points which were obscure in the
previous papers \cite{IMU2,IMU3}. 
We will discuss these applications in a separate paper \cite{IMU5}.

In appendix A, we summarize the relations between 
holomorphic and covariant HS currents up to spin 4.

\section{$\mathbf{U(1)}$ current and EM tensor }
\setcounter{equation}{0}
\label{sec 2}

In this section, we review a derivation of the holomorphic $U(1)$ 
and  EM tensor in the background of $U(1)$ gauge and
gravitational fields. These holomorphic quantities are obtained by
solving conservation equations together with the anomaly equations.

Throughout this paper we employ the conformal gauge $ds^2=e^{\varphi} du dv$
for the gravitational background and the Lorenz gauge $\nabla^\mu A_\mu=0$
for the gauge field background.

First we derive the holomorphic $U(1)$ current.
The $U(1)$ current $J^\mu$ satisfies the conservation equation
\begin{eqnarray}
 \label{cons-vec}
  \nabla_\mu J^\mu &=& 0,
\end{eqnarray}
and the chiral anomaly for the chiral current $J^{5 \mu}$ is
given by
\begin{equation}
 \label{cons-axial}
  \nabla_\mu J^{5\mu} = \frac{1}{2\pi}\epsilon^{\mu\nu}F_{\mu\nu}.
\end{equation}
Here the charge of the field is set $e=1$. $F_{\mu\nu}$ is the field strength
of the background gauge field and $\epsilon^{\mu\nu}$ is the covariant
antisymmetric tensor, $\epsilon^{uv}=2e^{-\varphi}=g^{uv}$. The chiral
current is related to the gauge current by 
$J^{5\mu} = \epsilon^{\mu\nu} J_\nu$.  
From eqs. (\ref{cons-vec}) and (\ref{cons-axial}), we find
\begin{equation}
 \partial_{v}
  \left( J_{u}-\frac{1}{\pi}A_u \right) = 0, \qquad 
  \partial_{u}
  \left(J_{v}-\frac{1}{\pi}A_{v} \right) = 0,
\end{equation}
where the gauge conditions are used. 
Hence we define the (anti-)holomorphic $U(1)$ currents as follows:
\begin{equation}
 j(u)\equiv J_{u}-\frac{1}{\pi}A_u, \qquad
  \tilde{j} (v) \equiv J_{v}-\frac{1}{\pi}A_{v}.
\label{U1-current}
\end{equation}
The holomorphic $U(1)$ currents
generate a combination of the holomorphic gauge 
transformation, which is a combination of  gauge and chiral
transformations.   
Note that these currents are not covariant under the $U(1)$ gauge
transformations.

Next we  derive the holomorphic EM tensor. The conservation
equation of the matter EM tensor is given by
\begin{eqnarray}
  \nabla_\mu {T^\mu}_\nu &=& F_{\mu\nu} J^\mu. 
   \label{cons-EM}
\end{eqnarray}
The r.h.s.   represents dissipation of the energy 
in the matter sector to the background gauge field. 
The trace anomaly of the EM tensor is given by   
\begin{eqnarray}
 {T^\mu}_\mu &=& \frac{c}{24\pi} R,
  \label{trace}
\end{eqnarray}
where $c$ is the  central charge of the matter field and $R$ is the Ricci
scalar $R=-4e^{-\varphi}\partial_u\partial_{v} \varphi$.
  From these equations, we obtain
\begin{align} 
&\partial_{v}
\left(  T_{uu}  - \frac{c}{24\pi}
  \left(\partial_u^2 \varphi - \frac{1}{2}(\partial_u \varphi)^2\right)
  - \frac{1}{\pi} A_u^2 - 2 A_u j(u)
  \right) =0.
\end{align} 
Thus we define the holomorphic energy-momentum tensor as
\begin{eqnarray}
t(u) \equiv
 T_{uu}  - \frac{c}{24\pi}
  \left(\partial_u^2 \varphi - \frac{1}{2}(\partial_u \varphi)^2\right)
  - \frac{1}{\pi} A_u^2 - 2 A_u j(u).
 \label{solveemtensor}
\end{eqnarray}
The anti-holomorphic one is defined similarly.  These currents
play a central role in conformal field theories since they generate
conformal transformations, which is a combination of 
the general coordinate, Weyl and chiral transformations.

\section{Holomorphic and covariant  HS currents}
\setcounter{equation}{0}
\label{sec 3}

In the previous section the relations between holomorphic and covariant
quantities are obtained in the cases of the $U(1)$ current and
energy-momentum tensor. In this section, we give a generalization of such
relations to higher-spin (HS) currents. We consider  fermionic fields in
the gravitational and electric backgrounds, and construct holomorphic
and covariant currents from them. Then we investigate the 
relations between these currents.

\subsection{Holomorphic HS currents}
In order to construct the holomorphic higher-spin
currents from  fermionic fields, let
us recall some properties of the fermion in the two dimensions. The equation
of motion for the right-handed fermion with unit charge is given by
\begin{align} 
\left( \partial_{v} -i A_{v} +\frac{1}{4}\partial_{v}\varphi
\right) \psi(u,{v})=0,
\label{equation-fermion}
\end{align} 
in the gravitational and electric backgrounds $(\varphi, A_\mu)$.  In the
Lorentz gauge, the gauge field can be written locally as
\begin{eqnarray}
 A_u = \partial_u \eta(u, v), \qquad
  A_v = -\partial_v \eta(u,v),
  \label{A-scalar}
\end{eqnarray}
where $\eta(u, v)$ is a scalar field.  Since gravitational fields and gauge
fields are not generally holomorphic, $\psi(u,v)$ is not
holomorphic either.  In order to construct holomorphic quantities from the
fermion field, we define a new field $\Psi$ as
\begin{align} 
\Psi \equiv 
\exp\left( \frac{1}{4}\varphi(u,{v}) +i \eta(u,v)
 \right) 
 \psi(u,{v}).
\label{Psi}
\end{align} 
Then the equation (\ref{equation-fermion}) becomes $\partial_{v}\Psi=0$ 
and hence $\Psi$ is holomorphic.
Similarly we can define $\Psi^\dagger$ as
\begin{align} 
\Psi^\dagger \equiv 
\exp\left( \frac{1}{4}\varphi(u,{v}) -i\eta(u,v)
 \right) 
 \psi^\dagger (u,{v}),
 \label{Psid}
\end{align} 
so that $\Psi^\dagger$ also becomes holomorphic. 

Regularized holomorphic currents are constructed from these holomorphic fields.
For example, the holomorphic $U(1)$ current can be defined as 
\begin{align} 
 j(u)=
 :\Psi^\dagger(u)\Psi(u):
 \equiv
 \lim_{\epsilon \rightarrow 0}
 \left[
 \Psi^\dagger \left( u+\frac{\epsilon}{2} \right) 
 \Psi\left( u-\frac{\epsilon}{2}\right) 
 +\frac{i}{2\pi \epsilon} 
 \right],
 \label{hol-U(1)curr}
\end{align} 
where the point splitting regularization is used and $\Psi$ has the following
operator product expansion,
\begin{equation} 
 \Psi^\dagger(u)\Psi(w) \sim
 -\frac{i}{2\pi} \frac{1}{u-w}.
 \label{ope}
\end{equation}
Note that we have not attached a Wilson line phase 
in the regularization, since the
gauge field is not holomorphic and the Wilson line phase 
breaks the holomorphy.
As a result, the holomorphic $U(1)$ current is not gauge invariant. We can
also construct  holomorphic currents $:\partial_u^n \Psi^\dagger
\partial_u^m \Psi(u) :$ in the same way.

In order to clarify the difference between the holomorphic current and the
ordinary covariant current, let us consider the covariant $U(1)$ current
$J_u$ in the electric background. We here omit the gravitational
background for simplicity.  $J_u$ can be defined as
 \begin{align}
  J_u \equiv
  \lim_{\epsilon \rightarrow 0} 
  \left[
  \psi^\dagger(u+\epsilon/2,{v}) 
  e^{i \int^{u+\epsilon/2}_{u-\epsilon/2} A_u (u',{v})du'} 
  \psi(u-\epsilon/2,{v})
  +\frac{i}{2\pi \epsilon} 
  \right] .
  \label{U(1)cov}
 \end{align}
In contrast with the holomorphic $U(1)$ current, we have attached the Wilson
line phase in the regularization, so this current is gauge invariant but
not holomorphic. By using (\ref{Psi}), (\ref{Psid}) and the operator product
expansion 
(\ref{ope}), the covariant $U(1)$ current can be  related to the holomorphic
$U(1)$ current (\ref{hol-U(1)curr}) as follows,
\begin{eqnarray} 
 J_u 
 &=& \lim_{\epsilon \rightarrow 0} 
  \left[
    e^{i \int^{u+\epsilon/2}_{u-\epsilon/2} A_u (u',{v})du'  
    +i \eta(u+\epsilon/2, v) -i \eta(u-\epsilon/2, v)} 
    \left(
     :\Psi^\dagger(u+\epsilon/2) \Psi(u-\epsilon/2):
     -\frac{i}{2\pi \epsilon}  
    \right) 
    + \frac{i}{2\pi \epsilon} 
   \right] \nonumber \\
 &=& j(u) +\frac{1}{\pi}A_u.
\label{hol-cov-U(1)curr}
\end{eqnarray}
This equation reproduces the equation
(\ref{U1-current}) which was originally derived from the conservation
equation (\ref{cons-vec}) and the chiral anomaly (\ref{cons-axial}). Hence
this evaluation of the covariant current is equivalent to solving the
conservation and anomaly equations.  Similarly the relations between the
covariant and the holomorphic HS currents contain the full information of
conservation equations and anomalies for these HS currents. We will discuss
it in the next section.

In the following subsections, we will consider a generalization of the
relation (\ref{hol-cov-U(1)curr}) to  the HS currents. 
Instead of considering each HS currents separately, 
it turns out that 
it is useful to introduce the following generating function of the
holomorphic currents
\begin{eqnarray}
 G_{hol}(u+a, u+b)
 &\equiv& \sum_{m, n=0}^\infty \frac{a^m b^n}{m!n!}
 :\partial_u^m \Psi^\dagger(u) \partial_u^n \Psi(u):
 \nonumber \\
 &=& \Psi^\dagger(u+a)\Psi(u+b)+\frac{i}{2\pi(a-b)}.
 \label{generating-holo}
\end{eqnarray}
This should be understood as a formal power series in terms of the
parameters $a$ and $b$ around the position $u$.
This function is holomorphic but not gauge covariant.
In the  subsection \ref{gen-fun-HS}, 
we will  construct a generating function for the
covariant currents,  and then give a relation between
these two functions. 
 
We here comment on the transformation property of the fermion field $\Psi$
under (holomorphic) gauge transformations. 
In the Lorentz gauge, there remains residual holomorphic gauge symmetry,
\begin{eqnarray}
 \psi'(u, v) = e^{i\Lambda(u)} \psi(u,v), \qquad
  \eta'(u,v) = \eta(u,v) + \Lambda(u).
\end{eqnarray} 
Under this transformation, $\Psi(u)$ transforms as a field with twice the
charge of $\psi$,
\begin{eqnarray}
 \Psi'(u) = e^{2i\Lambda(u)}\Psi(u).
\end{eqnarray} 
This clarifies a point which we did not explain explicitly in
\cite{IMU3}, i.e., we there used this transformation property for the
holomorphic field under the holomorphic transformation connecting a suitable
gauge at infinity and a suitable one near the horizon.

\subsection{Covariant HS currents}

Now we will define the $(u \cdots u)$-component of the covariant
 HS currents constructed from the
fermion $\psi$ in the electric and gravitational backgrounds.
Since these currents are covariant under holomorphic
general coordinate transformations, $u \rightarrow \tilde{u}= f(u)$,
it is convenient to define the coordinate which is invariant
under these transformations. 

In the rest of this section, we consider  $v$ to be a fixed coordinate 
and treat the system as  a one-dimensional one with the
coordinate $u$.
Then, under the  holomorphic general coordinate transformations,
we can define an ``invariant coordinate'' $x$ which satisfies
$\partial_x = e^{-\varphi}\partial_u$ and regard $u$ as a function of $x$,
i.e. $u=u(x)$.
Since $dx$ is invariant under the above holomorphic
transformations, the point splitting regularization is also
invariant if it is defined in the $x$ coordinate, 
not in the $u$ coordinate\footnote{
Since the coordinate $x$ is formally introduced as a function of 
$u$, $v$ should be kept fixed if a formula contains $x$ explicitly.
A derivative with respect 
to $v$ must be taken only after the $x$ coordinate
is removed.}.
$u(x+\epsilon)$ is now expanded  as a formal power series of $\epsilon$ as
\begin{eqnarray}
 u(x+\epsilon) &=& u(x) + \epsilon \partial_x u(x)
  + \frac{\epsilon^2}{2}\partial_x^2 u(x) + 
  \frac{\epsilon^3}{3!}\partial_x^3 u(x) + 
   \cdots 
  \nonumber \\
 &=& u(x) + \epsilon e^{-\varphi} \partial_u u(x) 
  + \frac{\epsilon^2}{2}\left(e^{-\varphi} \partial_u\right)^2 u(x)
    +  \frac{\epsilon^3}{3!} (e^{-\varphi} \partial_u)^3 +
\cdots 
  \nonumber \\
 &=& u + \epsilon e^{-\varphi} 
  -\frac{\epsilon^2}{2}e^{-2\varphi}\partial_u \varphi 
  -\frac{\epsilon^3}{6}e^{-3\varphi} 
   \left( \partial_u^2 \varphi - 2 (\partial_u \varphi)^2 
    \right)
  + \cdots,
\end{eqnarray}
where we used only the relation $\partial_x = e^{-\varphi}\partial_u$. It is
important that the last expression does not explicitly depend on $x$.
A field $\phi$ located at $u(x+\epsilon)$ is defined as the following
expansion,
\begin{eqnarray}
 \phi(u(x+\epsilon)) &=& \phi(u(x)) + \epsilon \partial_x \phi(u(x))
  + \frac{\epsilon^2}{2}\partial_x^2 \phi(u(x)) + \cdots
  \nonumber \\
 &=& \phi(u) + \epsilon e^{-\varphi} \partial_u \phi(u)
  + \frac{\epsilon^2}{2} \left(e^{-\varphi}\partial_u\right)^2 \phi(u)
  + \cdots.
\end{eqnarray}


Let's first 
consider the relation between the holomorphic and covariant
EM tensor in the electric and gravitational backgrounds.
As explained above, covariant regularization can be defined
in the $x$ coordinate as follows,
\begin{eqnarray}
 T_{uu} &\equiv& e^{2\varphi(u,v)}\lim_{\epsilon\rightarrow 0}
  \left\{
   -\frac{i}{2} e^{i\int^{u(x+\epsilon/2)}_{u(x-\epsilon/2)} du' A_u(u',v)}
   \right. \nonumber \\ 
 && \left[
     \left(e^{-\frac{5}{4}\varphi(u(x+\epsilon/2),v)}
      \nabla_u \psi^\dagger (u(x+\epsilon/2),v)\right)
     \left(e^{-\frac{1}{4}\varphi(u(x-\epsilon/2),v)}
      \psi (u(x-\epsilon/2),v)\right)
    \right.  \nonumber \\
 &&  \left. 
     -\left(e^{-\frac{1}{4}\varphi(u(x+\epsilon/2),v)}
       \psi^\dagger (u(x+\epsilon/2),v)\right)
     \left(e^{-\frac{5}{4}\varphi(u(x-\epsilon/2),v)}
      \nabla_u \psi (u(x-\epsilon/2),v)\right)
     \right] 
 \nonumber \\
 && \left.
     -\frac{1}{2\pi\epsilon^2}
  \right\}.
\end{eqnarray}
Here the Wilson line phase is introduced 
to guarantee the $U(1)$ gauge invariance. 
The covariant derivative is defined  by 
\begin{align}
\nabla_u \psi(u(x+\epsilon/2)) = \left(\partial_u -\frac{1}{4}\partial_u
\varphi(u(x+\epsilon/2)) -i A_u(u(x+\epsilon/2))\right)
\psi(u(x+\epsilon/2)).
\end{align}
Hence 
$e^{-\frac{5}{4}\varphi}\nabla_u \psi^\dagger$ and
$e^{-\frac{1}{4}\varphi}\psi^\dagger $ transform
as scalars 
under holomorphic coordinate transformations.
Therefore $T_{uu}$ transforms as a weight 2  tensor.

 This EM tensor can be 
rewritten in terms of the holomorphic fields by using eqs. (\ref{Psi}) and 
(\ref{Psid}) as
\begin{eqnarray}
 T_{uu} &=& e^{2\varphi(u,v)}\lim_{\epsilon\rightarrow 0}
  \left\{ -\frac{i}{2} 
   e^{2i\left(\eta(u(x+\epsilon/2)) - \eta(u(x-\epsilon/2))\right)}
   \right. \nonumber \\ 
 && \left[
      e^{-\frac{3}{2}\varphi(u(x+\epsilon/2),v)
      -\frac{1}{2}\varphi(u(x-\epsilon/2),v)}
      \nabla_u \Psi^\dagger (u(x+\epsilon/2),v)
      \Psi (u(x-\epsilon/2),v)
    \right.  \nonumber \\
 &&  \left. 
     -e^{-\frac{1}{2}\varphi(u(x+\epsilon/2),v)
     -\frac{3}{2}\varphi(u(x-\epsilon/2),v)}
       \Psi^\dagger (u(x+\epsilon/2),v)
      \nabla_u \Psi (u(x-\epsilon/2),v)
     \right] 
 \nonumber \\
 && \left.
     -\frac{1}{2\pi\epsilon^2}
  \right\},
\end{eqnarray}
where the covariant derivative of $\Psi$ is 
$\nabla_u \Psi = \left(\partial_u -\frac{1}{2}\partial_u \varphi -2i
A_u\right) \Psi$ and the gauge field is written by the scalar field $\eta$, 
eq. (\ref{A-scalar}). 
By using the following operator product expansion
\begin{eqnarray}
 \Psi^\dagger(u(x+\epsilon/2)) \Psi(u(x-\epsilon/2))
  +\frac{i}{2\pi}\frac{1}{u(x+\epsilon/2)-u(x-\epsilon/2)}
  = :\Psi^\dagger(u)\Psi(u):, 
  \label{ope-epsilon}
\end{eqnarray}
we find
\begin{eqnarray}
 T_{uu} &=& -\frac{i}{2}:\left(\partial_u \Psi^\dagger(u) \Psi(u) 
  - \Psi^\dagger(u) \partial_u \Psi(u)\right):
  + 2A_u(u,v) :\Psi^\dagger(u) \Psi(u):
  \nonumber \\
  && + \frac{1}{\pi}A_u^2(u,v) 
  + \frac{1}{24\pi}\left(\partial_u^2 \varphi(u,v) 
		    - \frac{1}{2}(\partial_u \varphi(u,v))^2\right).
\end{eqnarray}
This relation is equivalent to eq. (\ref{solveemtensor}) with the
following identification of the holomorphic EM tensor,
\begin{eqnarray}
 t(u) = -\frac{i}{2}:\left(\partial_u \Psi^\dagger(u) \Psi(u) 
  - \Psi^\dagger(u) \partial_u \Psi(u)\right):.
\end{eqnarray}

Generalizing these definitions of the covariant currents, we define the
covariant HS current $J^{(n)}_{u \cdots u}$ as follows,
\begin{eqnarray}
  J^{(n+1)}_{u\cdots u} &=& e^{(n+1)\varphi(u, v)}
  \lim_{\epsilon \rightarrow 0}
  \left[
   \sum_{m=0}^{n}\frac{n!}{2^n m! (n-m)!}
   e^{i\int^{u(x+\epsilon/2)}_{u(x-\epsilon/2)} du' A_u(u',v)}
  \right. \nonumber \\
 && \hspace{30mm}
  \times e^{-(n-m+1/4)\varphi(u(x+\epsilon/2), v)}
  \left(-i\nabla_u\right)^{n-m}\psi^\dagger (u(x+\epsilon/2), v)
  \nonumber \\
 && \hspace{30mm}
  \times e^{-(m+1/4)\varphi(u(x-\epsilon/2), v)}
  \left(i\nabla_u\right)^m\psi (u(x-\epsilon/2), v)
  \nonumber \\
 && \left. \hspace{30mm}
     + \frac{i^{n+1} n!}{2\pi \epsilon^{n+1}}
     \right].
 \label{def-cov-HS}
\end{eqnarray}
This current is symmetric with respect to $\psi$ and $\psi^\dagger$ and
invariant under the $U(1)$ gauge transformations thanks to the Wilson line
phase.  
The point splitting regularization is performed 
in the $x$ coordinate.
Furthermore we have multiplied the conformal factors at $u(x \pm
\epsilon/2)$ to make the combinations 
to be scalars under holomorphic general coordinate transformations.
Therefore, because of the factor $e^{(n+1)\varphi}$,
this current transforms as a tensor with a weight $(n+1)$
under holomorphic general coordinate transformations.
By
rewriting this quantity in terms of the holomorphic fields and using the
operator product expansion
(\ref{ope-epsilon}), we can take the limit $\epsilon \rightarrow 0$ and
get a formula which no more contains the formally introduced $x$
coordinate.  
Then a relation between the
covariant and the holomorphic HS currents is obtained.
The holomorphic HS currents are  defined
as\footnote{Our definitions of the HS currents are different
from those of the $W_{1+\infty}$ algebra in \cite{W}.  Their
HS currents are given by combining our HS currents and the derivative of them.}
\begin{eqnarray}
 j^{(n+1)} (u) 
  = \sum_{m=0}^{n}\frac{n!}{2^n m! (n-m)!}
  :\left(-i\partial_u\right)^{n-m} \Psi^\dagger 
  \left(i\partial_u\right)^m \Psi:.
  \label{def-hol-HS}
\end{eqnarray}
In these notations, we denote the U(1) currents and the EM tensors as
$J_u = J^{(1)}_u, T_{uu} = J^{(2)}_{uu}, j(u) = j^{(1)}(u)$ and 
$t(u) = j^{(2)}(u)$. 

The explicit forms of the relations between the covariant and holomorphic
currents with spin 3 and 4 will be given in section 4. 

\subsection{Generating functions of HS currents}
\label{gen-fun-HS}
Instead of studying each HS current separately, it is
 simpler and more systematic to consider a generating function of
the HS currents.
We  define the following
 generating function of the covariant HS currents as a
 formal power series with respect to a parameter $a$,
\begin{eqnarray}
 G_{cov}(a) &\equiv& \sum_{n=0}^{\infty} \frac{(2ia)^n}{n!}
  e^{-(n+1)\varphi(u,v)} J^{(n+1)}_{u\cdots u}. 
    \label{def-Gcov}
\end{eqnarray} 
In substituting the definition of $J^{(n+1)}_{u\cdots u}$, we use the
following relation,
\begin{eqnarray}
 && 
  e^{i\int^{u(x+\epsilon/2)}_{u_0} du' A_u(u',v)}
  e^{-(k+\frac{1}{4})\varphi(u(x+\epsilon/2))}
  \nabla_u^k \psi^\dagger (u(x+\epsilon/2)) 
  \nonumber \\
 && = 
  \left(e^{-\varphi(u(x+\epsilon/2))} 
   \frac{\partial}{\partial u(x+\epsilon/2)}\right)^k
  e^{i\int^{u(x+\epsilon/2)}_{u_0} du' A_u(u',v)}
  e^{-(k+\frac{1}{4})\varphi(u(x+\epsilon/2))}
  \psi^\dagger (u(x+\epsilon/2)), \nonumber \\
\end{eqnarray}
where $u_0$ is a fixed value on the $u$ coordinate.
A similar calculation can be done for the term including $\psi$.
Then $G_{cov}(a)$ can be represented as
\begin{eqnarray}
 G_{cov}(a) 
 &\hspace{-3mm}=\hspace{-3mm}& \lim_{\epsilon\rightarrow 0}
  \left[
   \sum_{m=0}^\infty \frac{a^m}{m!}
   \left(e^{-\varphi(u(x+\epsilon/2))} 
   \frac{\partial}{\partial u(x+\epsilon/2)} \right)^m
     e^{i\eta(u(x+\epsilon/2))-(k+\frac{1}{4})\varphi(u(x+\epsilon/2))}
   \psi^\dagger (u(x+\epsilon/2))
   \right. \nonumber\\
 && \hspace{0mm} 
  \times 
  \sum_{n=0}^\infty \frac{(-a)^n}{n!}
   \left(e^{-\varphi(u(x-\epsilon/2))} 
   \frac{\partial}{\partial u(x-\epsilon/2)}\right)^n
     e^{-i\eta(u(x-\epsilon/2))-(k+\frac{1}{4})\varphi(u(x-\epsilon/2))}
   \psi (u(x-\epsilon/2))
   \nonumber \\
 && \left. \hspace{10mm}
     +\frac{i}{4\pi(a+\epsilon/2)}
    \right]
 \nonumber \\
 &\hspace{-3mm}=\hspace{-3mm}& 
  e^{-\frac{1}{4}(\varphi(u(x+a),v) + \varphi(u(x-a),v))}
  e^{i\int^{u(x+a)}_{u(x-a)} du' A_u(u', v)}
  \psi^\dagger(u(x+a),v) \psi(u(x-a),v)
  \nonumber \\
 &&  + \frac{i}{4\pi a}.
  \label{Gcov}
\end{eqnarray}
In the last expression we have naively taken the limit
$\epsilon\rightarrow 0$ for notational simplicity, but the precise meaning
of $G_{cov}(a)$ is given by the first expression. By similar procedures to
those used in the previous subsection, i.e. employing eqs. (\ref{Psi}),
(\ref{Psid}) and (\ref{ope-epsilon}), the generating function can be
described in terms of the holomorphic fields,
\begin{eqnarray}
 G_{cov}(a) &=&  e^{-\frac{1}{2}(\varphi(u(x+a),v)+\varphi(u(x-a),v))}
  e^{2i \left(\eta(u(x+a),v) - \eta(u(x-a),v)\right)}   
  \nonumber \\
 && \times
  \left( G_{hol}(u(x+a), u(x-a))
   -\frac{i\hbar}{2\pi}\frac{1}{u(x+a)-u(x-a)}
  \right)
  \nonumber \\
  && +\frac{i\hbar}{4\pi a},
   \label{Gcov-hol}
\end{eqnarray}
where the gauge field is represented by using $\eta$, eq.(\ref{A-scalar}).
In this expression, in order to distinguish the quantum contributions from
the classical ones, we recovered the Planck's constant $\hbar$.  
  By expanding the relation with respect to
the parameter $a$, we can obtain equations which relate the covariant HS
currents with a sum of the holomorphic currents in the electric and
gravitational background.

We can also define a generating function for the holomorphic HS currents
$j^{(n)}(u)$, 
\begin{eqnarray}
 G_{hol}(u+\alpha, u-\alpha) &=& \sum_{n=0}^{\infty} \frac{(2i\alpha)^n}{n!} 
  j^{(n+1)}(u) 
  \label{def-Ghol}
\end{eqnarray}
By using eqs.(\ref{Psi}) and (\ref{Psid}), $G_{hol}(\alpha)$ can be written
in terms of $\psi$ and $\psi^\dagger$ as
\begin{eqnarray}
 G_{hol}(u+\alpha, u-\alpha) 
  &=& e^{\frac{1}{2}(\varphi(u+\alpha)+\varphi(u-\alpha))
  -2i \left(\eta(u+\alpha) - \eta(u-\alpha)\right)} 
  \nonumber \\
 && \times\Biggl[
  e^{-\frac{1}{4}(\varphi(u+\alpha)+\varphi(u-\alpha))
  +i \left(\eta(u+\alpha) - \eta(u-\alpha)\right)}
  \psi^\dagger(u+\alpha, v)\psi(u-\alpha, v) 
  \nonumber \\
 && +\frac{i\hbar}{2\pi}\frac{1}{x(u+\alpha)-x(u-\alpha)}   \Biggr]
  -\frac{i\hbar}{2\pi}
  \frac{ e^{\frac{1}{2}(\varphi(u+\alpha)+\varphi(u-\alpha))
  -2i \left(\eta(u+\alpha) - \eta(u-\alpha)\right)  }}
  {x(u+\alpha)-x(u-\alpha)}
  \nonumber \\
  && +\frac{i\hbar}{4\pi \alpha}. 
\label{Ghol-cov}
\end{eqnarray}
The first term in (\ref{Ghol-cov}) can be described in terms of the
covariant HS currents and their derivatives. By expanding (\ref{Ghol-cov})
with respect to the parameter $\alpha$, the equations relating the
holomorphic HS currents with a sum of the covariant ones can be
derived. The obtained relations are summarized in the appendix
\ref{App-current}.
 
In the next section, we investigate the relations for spin 3 and 4
currents, and discuss the conservation equations, trace anomalies and
generalizations of gauge (or gravitational) anomalies.

\section{Trace anomalies for HS currents}
\setcounter{equation}{0}
\label{sec 4}

In section \ref{sec 2}, we derived the relation between $T_{uu}$ and $t(u)$
from the conservation equation
(\ref{cons-EM}) and the anomaly equation (\ref{trace}).
In this section, we take an inverse step for the HS currents.
We first provide  relations between the holomorphic and
covariant HS currents from equation (\ref{Gcov-hol}), and then, by
using these relations,  evaluate their conservation equations and
trace anomalies. 
In order to fix the definitions of the currents, we
impose the following assumptions about the currents:
\begin{enumerate}
  \item Anomalies appear in the trace parts of the currents only.
  \item The covariant currents are classically traceless.
  \item The covariant currents are totally symmetric.
\end{enumerate}
Under these assumptions, we will obtain the conservation equations and trace
anomalies for the spin 3 and 4 currents. 
The derivation can be straightforwardly
applied to  general HS currents, though
calculations become more complicated.

We use the following notations of the currents in this section.
$J^{(1)}_\mu$ denotes the covariant $U(1)$ current $J_\mu$,
$J^{(2)}_{\mu\nu}$ denotes the covariant energy-momentum tensor $T_{\mu\nu}$
and $J^{(n)}_{\mu_1 \dots \mu_n}$ does the spin $n$ covariant current.

\subsection{Trace anomaly for spin 3 current}
We here consider the spin $3$ current.
From eq.(\ref{def-cov-HS}), the covariant spin $3$ current is given by
\begin{eqnarray}
J^{(3)}_{uuu} &=& -\frac{1}{4}e^{3\varphi} \lim_{\epsilon\rightarrow 0}
 e^{i \int^{u(x+\epsilon/2)}_{u(x-\epsilon/2)} du' A_u(u',v)}
 \nonumber \\
 && \hspace{-5mm} \times
 \left[
  e^{-\frac{9}{4}\varphi(u(x+\epsilon/2),v)
  -\frac{1}{4}\varphi(u(x-\epsilon/2),v)}
  \nabla_u^2 \psi^\dagger(u(x+\epsilon/2),v) \psi(u(x-\epsilon/2),v)
  \right. \nonumber \\
 && -2 e^{-\frac{5}{4}\varphi(u(x+\epsilon/2),v)
  -\frac{5}{4}\varphi(u(x-\epsilon/2),v)}
  \nabla_u \psi^\dagger(u(x+\epsilon/2),v) \nabla_u \psi(u(x-\epsilon/2),v)
  \nonumber \\
 && \left.
     + e^{-\frac{1}{4}\varphi(u(x+\epsilon/2),v)
     -\frac{9}{4}\varphi(u(x-\epsilon/2),v)}
     \psi^\dagger(u(x+\epsilon/2),v) \nabla_u^2 \psi(u(x-\epsilon/2),v)
    \right].
\end{eqnarray}
The corresponding holomorphic spin $3$ current is
\begin{align} 
j^{(3)}(u) \equiv -\frac{1}{4}
\left( 
 :\Psi^\dagger \partial_u^2 \Psi -2 \partial_u \Psi^\dagger \partial_u \Psi+
 \partial_u^2 \Psi^\dagger \Psi
 :\right).
\end{align} 
The following relation between these currents can be derived by
expanding $G_{cov}(a)$ (\ref{Gcov-hol}) and taking the $a^2$ terms as
\begin{align}
 J^{(3)}_{uuu} =& j^{(3)}(u)+4 A_u j^{(2)}(u)
  +\left(\frac{1}{4} 
    \left( \partial_u^2 \varphi - (\partial_u \varphi)^2 \right) 
    +4 A_u^2 \right) j^{(1)}(u) 
  +\frac{1}{4} \partial_u \varphi \partial_u j^{(1)}(u) 
  \nonumber \\
 & +\frac{\hbar}{4\pi}\left( A_u
  \left(\partial_u^2\varphi - (\partial_u \varphi)^2\right)
  + \partial_u \varphi \partial_u A_u
  -\frac{1}{3} \partial_u^2 A_u
  + \frac{16}{3}A_u^3\right) .
  \label{rank3curr}
\end{align}
From this equation, we derive the conservation equation and anomaly
equation for the covariant spin $3$ current.  First let us consider the $uu$
component of the conservation equations $\nabla_\mu {J^{(3)\mu}}_{uu}$.  By
taking the derivative of eq. (\ref{rank3curr}) with respect to $v$, we find
\begin{align} 
 \nabla_v J^{(3)}_{uuu}
 &=-2 F_{uv} J^{(2)}_{uu}
 -\frac{1}{8}\nabla_u\left( g_{uv}R  J^{(1)}_u \right) 
 +\frac{\hbar}{24\pi} \nabla_u^2 F_{uv}. 
\label{divergence spin 3}
\end{align} 
Here we have used the equations in appendix \ref{App-current} to describe
the holomorphic currents in terms of the covariant currents.  
Since, as mentioned above, we assume that anomalies arise only in the trace
part of the currents, we regard the last term in eq. 
(\ref{divergence spin 3}), which is a quantum contribution, 
as the covariant derivative of the trace anomaly,
\begin{align} 
 \nabla_u J^{(3)}_{vuu} = -\frac{\hbar}{24\pi} \nabla_u^2 F_{uv}.
\label{Jvuu}
\end{align} 
Thus the $uu$ component of the conservation equation becomes
\begin{align} 
 \nabla_\mu {J^{(3)\mu}}_{uu}
 &=-2 g^{uv} F_{uv} J^{(2)}_{uu}
 -\frac{1}{8}\nabla_u\left(R  J^{(1)}_u \right) . 
 \label{conservation uu}
\end{align} 
Here we have multiplied (\ref{divergence spin 3}) and (\ref{Jvuu}) by
$g^{uv}$. From this equation we may naively guess the general components of
the conservation equation as follows,
\begin{align} 
 \nabla_\mu {J^{(3)\mu}}_{\nu\rho}
 &=-  F_{\nu\mu} {J^{(2)\mu}}_\rho-  F_{\rho\mu} {J^{(2)\mu}}_{\nu}
 -\frac{1}{16}\nabla_\nu\left(R  J^{(1)}_\rho \right)  
 -\frac{1}{16}\nabla_\rho\left(R  J^{(1)}_\nu \right), 
\end{align} 
where the indices $\nu, \rho$ are symmetrized.  But this is not traceless
at the classical level, which contradicts with the second assumption.
Note that 
terms proportional to $g_{\nu \rho}$ can be added to the
conservation law without affecting eq. (\ref{conservation uu}). Thus by
using this freedom, we can make the r.h.s. of 
the conservation equation traceless with
respect to $\nu$ and $\rho$,
\begin{align} 
 \nabla_\mu {J^{(3)\mu}}_{\nu\rho}
 &=-  F_{\nu\mu} {J^{(2)\mu}}_\rho-  F_{\rho\mu} {J^{(2)\mu}}_{\nu}
 -\frac{1}{16}\nabla_\nu\left(R  J^{(1)}_\rho \right)  
 -\frac{1}{16}\nabla_\rho\left(R  J^{(1)}_\nu \right)
 + \frac{1}{16}g_{\nu\rho}\nabla_\mu \left(R J^{(1)\mu} \right) .
 \label{conservation 3}
\end{align} 
This satisfies the three conditions we require.

Next the trace anomaly $J^{(3)}_{vuu}$ can be read from equation (\ref{Jvuu}),
\begin{align} 
J^{(3)}_{vuu} = -\frac{\hbar}{24\pi} \nabla_u F_{uv}.
\end{align} 
This can be covariantized as
\begin{align} 
{J^{(3)\mu}}_{\mu \nu} = \frac{\hbar}{12\pi} \nabla_\mu {F^\mu}_{\nu}.
\label{anomaly 3}
\end{align} 
In order to check the consistency with the conservation law
(\ref{conservation 3}), we calculate $J^{(3)}_{uvv}$. Since $J^{(3)}_{uvv}$
is given by
\begin{align} 
J^{(3)}_{uvv} = -\frac{\hbar}{24\pi} \nabla_v F_{vu},
\end{align} 
we can show $\nabla_\mu {J^{(3)\mu}}_{uv}=0$ by using the identity in two
dimensions, $[\nabla_\mu ,\nabla_\nu] F_{\rho \sigma}=0$. Hence
(\ref{anomaly 3}) is consistent with (\ref{conservation 3}).

The conservation equation (\ref{conservation 3}) implies that the theory can
possess  symmetry associated with the spin 3 current if the corresponding
spin 3 gauge field is included in it. Let us consider an action containing
linear couplings of the HS currents to general higher-spin gauge fields
$B^{(n)}_{\mu_1 \cdots \mu_n}$,
\begin{eqnarray}
 S[A, g, B^{(n)}] = \int d^2x \sqrt{-g}
  \left(
   {\cal L}_0 + \sum_{n=3}^{\infty} \frac{1}{n!} 
   B^{(n)}_{\mu_1 \cdots \mu_n}J^{(n) \mu_1 \cdots \mu_n}
  \right),
\end{eqnarray}
where ${\cal L}_0$ is the Lagrangian for the free fermion in the
electric and gravitational backgrounds, $A_\mu$ and $g_{\mu\nu}$,
respectively. We also introduce the following effective action for these
gauge fields,
\begin{eqnarray}
 e^{i\Gamma[A, g, B^{(n)}]} = \int {\cal D}\bar{\psi}{\cal D}\psi~
  e^{iS[A, g, B^{(n)}]}.
\end{eqnarray} 
The expectation values of the HS currents in these backgrounds are
given by
\begin{eqnarray}
 && \langle J^{(1)}_\mu \rangle = \langle J_\mu \rangle 
  = \frac{1}{\sqrt{-g}}\frac{\delta}{\delta A^\mu}  \Gamma[A, g, B^{(n)}], 
  \nonumber \\
 && \langle J^{(2)}_{\mu\nu} \rangle = \langle T_{\mu\nu} \rangle
  = \frac{2}{\sqrt{-g}}\frac{\delta}{\delta g^{\mu\nu}} 
  \Gamma[A, g, B^{(n)}],
  \nonumber \\
 && \langle J^{(n)}_{\mu_1 \cdots \mu_n} \rangle 
  = \frac{1}{\sqrt{-g}}\frac{\delta}{\delta B^{(n)\mu_1 \cdots \mu}_n} 
  \Gamma[A, g, B^{(n)}], \qquad
  (n \geq 3).
\end{eqnarray}
Then the conservation equation (\ref{conservation 3}) indicates that the
effective action is invariant under the following infinitesimal
transformations  of the background  fields,
\begin{eqnarray}
 \delta_{\xi} B^{(3) \mu\nu\rho} 
  &=& \frac{1}{3} 
  \left(\nabla^\mu \xi^{\nu\rho} + \nabla^\nu \xi^{\rho\mu} 
   + \nabla^\rho \xi^{\mu\nu} 
  \right) 
  \label{transformation rank 3}\\
 \delta_{\xi}g^{\mu\nu} 
 &=& -2\xi^{\mu\sigma}{F_{\sigma}}^\nu -2\xi^{\nu\sigma}{F_{\sigma}}^\mu,\\
 \delta_{\xi}A^\mu 
  &=& \frac{1}{8}R \nabla_\nu \xi^{\nu \mu},
  \label{transformation rank 1} 
\end{eqnarray}
where $\xi^{\mu\nu}$ is a symmetric traceless parameter. 

This transformation law is valid only 
for the weak $B^{(3)}$ field limit, i.e.
 we have assumed that  
 the rank 3 gauge field $B^{(3)}$ was originally 
absent.   Since the OPE between spin 3 currents
generate higher-spin currents, they 
no longer form a closed algebra, contrary to the spin 1 or
spin 2 currents. 
Hence higher-spin gauge symmetries larger than 2
and their backgrounds 
must be considered  as $W_{\infty}$ gauge symmetry
and gauge fields as a whole. 
This is beyond the scope of the present paper.

\subsection{Trace anomaly for spin 4 current}

The covariant spin 4 current is given from eq. (\ref{def-cov-HS}) by
\begin{eqnarray}
J^{(4)}_{uuu} &=& \frac{i}{8} e^{4\varphi} \lim_{\epsilon\rightarrow 0}
 e^{i \int^{u(x+\epsilon/2)}_{u(x-\epsilon/2)} du' A_u(u',v)}
 \nonumber \\
 && \hspace{-5mm} \times
 \left[
  e^{-\frac{13}{4}\varphi(u(x+\epsilon/2),v)
  -\frac{1}{4}\varphi(u(x-\epsilon/2),v)}
  \nabla_u^3 \psi^\dagger(u(x+\epsilon/2),v) \psi(u(x-\epsilon/2),v)
  \right. \nonumber \\
 && -3 e^{-\frac{9}{4}\varphi(u(x+\epsilon/2),v)
  -\frac{5}{4}\varphi(u(x-\epsilon/2),v)}
  \nabla_u^2 \psi^\dagger(u(x+\epsilon/2),v) \nabla_u\psi(u(x-\epsilon/2),v) 
  \nonumber \\
 && +3 e^{-\frac{5}{4}\varphi(u(x+\epsilon/2),v)
  -\frac{9}{4}\varphi(u(x-\epsilon/2),v)}
  \nabla_u\psi^\dagger(u(x+\epsilon/2),v) \nabla_u^2 \psi(u(x-\epsilon/2),v)
  \nonumber \\
 && \left.
  - e^{-\frac{1}{4}\varphi(u(x+\epsilon/2),v)
  -\frac{13}{4}\varphi(u(x-\epsilon/2),v)}
  \psi^\dagger(u(x+\epsilon/2),v) \nabla_u^3 \psi(u(x-\epsilon/2),v)
  \right],
\end{eqnarray}
and the corresponding holomorphic current is
\begin{align} 
 j^{(4)}(u)
 =& \frac{i}{8}:\partial_u^3 \Psi^\dagger \Psi 
 -3 \partial_u^2 \Psi^\dagger \partial_u \Psi
 +3 \partial_u \Psi^\dagger \partial_u^2 \Psi
 - \Psi^\dagger \partial_u^3 \Psi:.
\end{align} 
The relation between these two currents are obtained from
$a^3$ terms of the equation (\ref{Gcov-hol}) as
\begin{align}
 J^{(4)}_{uuuu}
  =& j^{(4)}(u)  
  +6A_u j^{(3)}(u) +\frac{3}{4}\partial_u \varphi \partial_u j^{(2)}(u)
  + \left[
     \frac{1}{4}(4\partial_u^2 \varphi - 5 (\partial_u \varphi)^2)
     +12 A_u^2
    \right] j^{(2)}(u)
  \nonumber \\
 & +\frac{3}{2} A_u \partial_u \varphi \partial_u j^{(1)}(u)
  + \left[
     2 A_u \left(\partial_u^2 \varphi 
	    - \frac{5}{4} (\partial_u \varphi)^2\right)
     + \frac{3}{2}\partial_u A_u \partial_u \varphi
     -\frac{1}{2} \partial_u^2 A_u 
     + 8 A_u^3
    \right] j^{(1)}(u)
  \nonumber \\
 & - \frac{\hbar}{2\pi} A_u 
  (\partial_u - 2 \partial_u \varphi)(\partial_u - \partial_u \varphi)A_u
  + \frac{\hbar}{2\pi} A_u^2
  \left(\partial_u^2 \varphi - \frac{1}{2}(\partial_u \varphi)^2\right)
  + \frac{2\hbar}{\pi} A_u^4
  \nonumber \\
 & - \frac{\hbar}{160\pi}(\partial_u - 3 \partial_u \varphi)
  (\partial_u - 2 \partial_u \varphi)
  \left(\partial_u^2 \varphi - \frac{1}{2}(\partial_u \varphi)^2\right)
  + \frac{7\hbar}{480\pi}
  \left(\partial_u^2 \varphi - \frac{1}{2}(\partial_u \varphi)^2\right)^2.
  \label{spin4-hol-cov}
\end{align}
First we derive the $uuu$ component of the conservation equation from this
equation by taking the derivative of (\ref{spin4-hol-cov}) with respect
to $v$ and multiplying $g^{uv}$,
\begin{align} 
 \nabla^u J^{(4)}_{uuuu}=& 3 g^{uv} F_{vu} J^{(3)}_{uuu}
 -\frac{3}{8}R \nabla_u J^{(2)}_{uu}-\frac{1}{2}J^{(2)}_{uu} \nabla_u R
 -\frac{1}{4}g^{uv}J^{(1)}_{u}\nabla_u^2F_{vu}+ \frac{\hbar}{320\pi}\nabla_u^3 R .
 \label{divergence spin 4}
\end{align} 
As in the case of the spin $3$ current, we regard the last term as the
contribution of the trace anomaly because it is proportional to $\hbar$
and quantum. 
From the
assumptions 1 and 2, $J^{(4)}_{vuuu}$ is given by
\begin{align} 
 J^{(4)}_{vuuu}=-\frac{\hbar}{320\pi}g_{uv}\nabla_u^2R.
\label{Jvuuu}
\end{align} 
{}From (\ref{divergence spin 4}), we guess the covariant conservation
equation as
\begin{align} 
 \nabla^\mu J^{(4)}_{\mu\nu\rho\sigma}=& 
 F_{\mu\nu} {J^{(3)\mu}}_{\rho \sigma}
 + F_{\mu\rho} {J^{(3)\mu}}_{ \sigma\nu}
 +F_{\mu\sigma} {J^{(3)\mu}}_{\nu\rho } 
 -\frac{1}{8}R
 \left(  
 \nabla_\nu J^{(2)}_{\rho\sigma} 
 + \nabla_\rho J^{(2)}_{\sigma\nu} 
 +  \nabla_\sigma J^{(2)}_{\nu\rho} 
 \right) \nonumber \\
 &-\frac{1}{6}
 \left( 
 J^{(2)}_{\nu\rho} \nabla_\sigma R 
 +J^{(2)}_{\rho\sigma} \nabla_\nu R
 + J^{(2)}_{\sigma\nu} \nabla_\rho R 
 \right) \nonumber \\
 &-\frac{1}{24} \Biggl( J^{(1)}_{\nu}\nabla_\rho \nabla_\mu {F^\mu}_{\sigma} 
 + J^{(1)}_{\rho}\nabla_\sigma \nabla_\mu {F^\mu}_{\nu}
 + J^{(1)}_{\sigma}\nabla_\nu \nabla_\mu {F^\mu}_{\rho} \nonumber \\
 &+J^{(1)}_{\rho}\nabla_\nu \nabla_\mu {F^\mu}_{\sigma}
 + J^{(1)}_{\nu}\nabla_\sigma \nabla_\mu {F^\mu}_{\rho}
 + J^{(1)}_{\sigma}\nabla_\rho \nabla_\mu {F^\mu}_{\nu}  \Biggr).
 \label{conservation spin 4}
\end{align} 
Next we add appropriate terms proportional to $g_{\mu\nu}$ so that this
conservation equation becomes classically traceless. In general, one
can construct a rank 3 traceless symmetric tensor from any rank 3 symmetric
tensor $B_{\nu\rho\sigma}$ by subtracting the trace part $(g_{\nu\rho}
{B^\mu}_{\mu\sigma}+g_{\sigma\rho} {B^\mu}_{\mu\nu}+g_{\sigma\nu}
{B^\mu}_{\mu\rho})/4$.  We define $C_{\nu}$ as the trace of
(\ref{conservation spin 4}),
\begin{eqnarray} 
 C_{\nu} &\equiv&  g^{\rho\sigma}\nabla^\mu J^{(4)}_{\mu\nu\rho\sigma}
  \nonumber \\
  &=& F_{\mu\nu} {J^{(3)\mu\rho}}_{\rho }
  -\frac{1}{8}R
  \nabla_\nu {J^{(2)\rho}}_\rho 
  -\frac{1}{4}R \nabla_\rho {J^{(2)\rho}}_{\nu} 
  -\frac{1}{3} {J^{(2)\rho}}_{\nu} \nabla_\rho R 
  -\frac{1}{6}{J^{(2)\rho}}_{\rho} \nabla_\nu R  
  \nonumber \\
 && -\frac{1}{12} \Biggl(  J^{(1)\rho}\nabla_\rho \nabla_\mu {F^\mu}_{\nu}
  +J^{(1)}_{\rho}\nabla_\nu \nabla_\mu F^{\mu\rho}  \Biggr).
\end{eqnarray}
Note that $C_\nu$ includes the traces of the covariant spin 2 and 3 currents
which vanish classically but not at the quantum level due to the trace
anomalies.  According to our assumption 1, we treat such
anomalous quantities as contributions of the trace anomaly. Therefore we
define $\tilde{C}_{\nu}$ as $C_{\nu}$ without the anomalous terms,
\begin{align} 
 \tilde{C}_{\nu}\equiv&  -\frac{1}{4}R \nabla_\rho {J^{(2)\rho}}_{\nu} 
 -\frac{1}{3} {J^{(2)\rho}}_{\nu} \nabla_\rho R 
 -\frac{1}{12} \Biggl(  J^{(1)\rho}\nabla_\rho \nabla_\mu {F^\mu}_{\nu}
 +J^{(1)}_{\rho}\nabla_\nu \nabla_\mu F^{\mu\rho}  \Biggr),
\end{align} 
and construct a new conservation equation, which are classically
traceless,
\begin{eqnarray}
 \nabla^\mu J^{(4)}_{\mu\nu\rho\sigma} &=&  
  F_{\mu\nu} {J^{(3)\mu}}_{\rho \sigma}
  + F_{\mu\rho} {J^{(3)\mu}}_{ \sigma\nu}
  + F_{\mu\sigma} {J^{(3)\mu}}_{\nu\rho } 
  -\frac{1}{8}R
  \left(  
   \nabla_\nu J^{(2)}_{\rho\sigma} 
   + \nabla_\rho J^{(2)}_{\sigma\nu}
   + \nabla_\sigma J^{(2)}_{\nu\rho} 
  \right) 
  \nonumber \\
 && -\frac{1}{6}
  \left( 
   J^{(2)}_{\nu\rho} \nabla_\sigma R 
   +J^{(2)}_{\rho\sigma} \nabla_\nu R
   + J^{(2)}_{\sigma\nu} \nabla_\rho R 
  \right) 
  \nonumber \\
 && -\frac{1}{24} 
  \Biggl( 
  J^{(1)}_{\nu}\nabla_\rho \nabla_\mu {F^\mu}_{\sigma} 
  +J^{(1)}_{\rho}\nabla_\sigma \nabla_\mu {F^\mu}_{\nu}
  +J^{(1)}_{\sigma}\nabla_\nu \nabla_\mu {F^\mu}_{\rho} 
  \nonumber \\
 && +J^{(1)}_{\rho}\nabla_\nu \nabla_\mu {F^\mu}_{\sigma}
  +J^{(1)}_{\nu}\nabla_\sigma \nabla_\mu {F^\mu}_{\rho}
  +J^{(1)}_{\sigma}\nabla_\rho \nabla_\mu {F^\mu}_{\nu}  
  \Biggr) 
  \nonumber \\
 && -\frac{1}{4}
  \left(
   g_{\nu\rho}\tilde{C}_{\sigma}+g_{\rho\sigma}\tilde{C}_{\nu}
   +g_{\sigma\nu}\tilde{C}_{\rho}  \right) .
  \label{conservation covariant spin 4}
\end{eqnarray}

Next the $uu$ component of the trace anomaly can be read from (\ref{Jvuuu}), 
\begin{align} 
 {J^{(4)\mu }}_{\mu uu}=-\frac{\hbar}{160\pi}\nabla_u^2R.
 \label{trace uu}
\end{align} 
Then general components of the trace anomaly have the following form,
\begin{align} 
 {J^{(4)\mu }}_{\mu \nu\rho}=-\frac{\hbar}{160\pi} \nabla_\nu \nabla_\rho R
 +g_{\nu\rho}A,
 \label{trace anomaly A}
\end{align} 
where $A$ is not fixed from (\ref{trace uu})
only.  We can determine $A$ by imposing consistency of (\ref{trace anomaly A}) 
with (\ref{conservation covariant spin 4}). The trace of
(\ref{conservation covariant spin 4}) becomes
\begin{align} 
 \nabla_\mu{J^{(4)\mu \rho}}_{\rho \nu}=&
 F_{\mu\nu} {J^{(3)\mu\rho}}_{\rho }
 -\frac{1}{8}R  \nabla_\nu {J^{(2)\rho}}_\rho  
 -\frac{1}{6}{J^{(2)\rho}}_{\rho} \nabla_\nu R  \nonumber \\
 =&\frac{\hbar}{24\pi} \nabla_\rho
 \left( {\tilde{F}}^2-\frac{7}{48}R^2 \right), 
\end{align} 
where $\tilde{F}\equiv \epsilon^{\mu\nu}F_{\mu\nu}/2=g^{uv}F_{uv}$.
On the other hand, the divergence of the (\ref{trace anomaly A}) is
 \begin{align} 
  \nabla_\mu{J^{(4)\mu \rho}}_{\rho \nu}=
  &-\frac{\hbar}{160\pi}
  \left(\nabla_\rho \nabla^2 R + \frac{1}{4} \nabla_\rho R^2 \right)
  + \nabla_\rho A .
 \end{align} 
 By comparing these two equations, $A$ is determined as
 \begin{align} 
  A=\frac{\hbar}{160\pi}\nabla^2 R
  +\frac{\hbar}{24\pi}\left({\tilde{F}}^2-\frac{13}{120}R^2  \right). 
 \end{align} 
As a result, we obtain the trace anomaly of the spin 4 current,
 \begin{align} 
 {J^{(4)\mu }}_{\mu \nu\rho}
  =-\frac{\hbar}{160\pi} \nabla_\nu \nabla_\rho R
  +g_{\nu\rho}\left[
  \frac{\hbar}{160\pi}\nabla^2 R+\frac{\hbar}{24\pi}\left({\tilde{F}}^2
  -\frac{13}{120}R^2  \right)  \right] .
 \end{align} 
 
As in the case of the rank 3 current, we can evaluate the transformation of
the background fields from the conservation equation 
(\ref{conservation covariant spin 4}),
\begin{eqnarray}
 \delta_{\xi} B^{(4) \mu\nu\rho\sigma}   
 &=&  \frac{1}{4}
 \left(\nabla^\mu \xi^{\nu\rho\sigma} + 
 \nabla^\nu \xi^{\rho\sigma\mu} + \nabla^\rho \xi^{\sigma\mu\nu}
 + \nabla^\sigma \xi^{\mu\nu\rho}\right)\\
 \delta_{\xi} B^{(3) \mu\nu\rho}
 &=& -3 \xi^{\mu\nu \sigma} {F_{\sigma}}^{\rho}
 -3 \xi^{\nu\rho \sigma} {F_{\sigma}}^{\mu}
 -3 \xi^{\rho\mu \sigma} {F_{\sigma}}^{\nu}, \\
 \delta_{\xi} g^{\mu\nu}
 &=&
 \frac{3}{4} \nabla_\rho \left(\xi^{\rho\mu\nu}R   \right)
 -\xi^{\mu\nu\rho}\nabla_\rho R 
 -\frac{3}{16} \nabla^\mu \left( R {\xi_\rho}^{\rho \nu} \right) 
 -\frac{3}{16} \nabla^\nu \left( R {\xi_\rho}^{\rho \mu} \right) 
 \nonumber \\
 && +\frac{1}{4} {\xi_\rho}^{\rho \mu} \nabla^\nu  R  
  +\frac{1}{4} {\xi_\rho}^{\rho \nu} \nabla^\mu  R  \\
 \delta_{\xi} A^{\mu}
 &=&
 -\frac{1}{4}\xi^{\mu\rho\sigma}\nabla_\rho \nabla_\nu {F^{\nu}}_{\sigma}
 +\frac{1}{16}
 {\xi_\rho}^{\rho \sigma}
 \left[\nabla^\mu \nabla_\nu {F^{\nu}}_{\sigma}
 +\nabla_\sigma \nabla_\nu F^{\nu\mu}
 \right],
\end{eqnarray}
where $\xi^{\mu\nu\rho}$ denotes a symmetric traceless parameter.
 
\section{Higher-spin gauge anomalies}
\setcounter{equation}{0} 
\label{sec chiral} 
In the previous section, we have obtained 
the conservation equations and trace anomalies 
in the HS currents by considering non-chiral
theories; i.e. the anomaly coefficients are the same
between the holomorphic and anti-holomorphic sectors.

In this section, we consider a chiral fermionic theory where we have
 $c_L$ left-handed fermions and $c_R(\neq c_L)$
 right-handed fermions.
In this case, the conservation equation  becomes anomalous.
This is  a generalization of the gauge or gravitational anomalies
to the HS currents. 
If these HS currents are coupled to HS gauge fields,
these violation of conservation equations 
lead to quantum violation of HS local symmetries. 

We here remark that, in the presence of $c_R$ right-handed 
and $c_L$ left-handed fermions,
the coefficients of the anomalous terms in the
$(u \cdots u)$ sector are multiplied by $c_R$,
and those in $(v \cdots v)$ sector by $c_L$.

In the following  of this section, we will derive the anomalous 
conservation equations  for the  currents up to rank 4.

\subsection{ $U(1)$ gauge and gravitational anomalies}
In this subsection, we reproduce the gauge and gravitational anomalies
from the relations between the (anti-) holomorphic and covariant $U(1)$ and 
spin 2 currents.

First we consider the $U(1)$ current. 
The relations in the present case with $c_L\ne c_R$ become
\begin{align}
 J_{u}=j(u)+\frac{c_R \hbar}{\pi}A_u, \qquad
  J_{v}= \tilde{j} (v) +\frac{c_L \hbar}{\pi}A_{v}.
\end{align}
By taking  derivatives of these equations, we obtain
\begin{align}
 \nabla_v J_{u}
 + \nabla_u J_{v}
& = \frac{(c_R-c_L)}{2}\frac{ \hbar}{\pi}  F_{vu}, \\
 \nabla_v J_{u}
 - \nabla_u J_{v}
& = \frac{(c_R+c_L)}{2}\frac{ \hbar}{\pi}  F_{vu},
\end{align}
where we have used the Lorenz gauge condition 
$\partial_u A_v= -\partial_v A_u$.
They can  be written  in the covariant forms as
\begin{align}
 \nabla_\mu J^{\mu}
& = -\frac{(c_R-c_L)}{2}\frac{ \hbar}{2\pi} \epsilon^{\mu\nu}  F_{\mu\nu} ,\\
 \nabla_\mu J^{5\mu}
& = \frac{(c_R+c_L)}{2}\frac{ \hbar}{2\pi} \epsilon^{\mu\nu}  F_{\mu\nu}.
\end{align}
Thus, if $c_L \ne c_R$, the gauge symmetry is broken by the anomaly.

Next we consider the energy-momentum tensor.
Now the relation (\ref{solveemtensor}) is modified as
\begin{eqnarray}
t(u) =
 T_{uu}   - 2 A_u j(u)  - \frac{c_R \hbar}{\pi} A_u^2
- \frac{c_R\hbar}{24\pi}
  \left(\partial_u^2 \varphi - \frac{1}{2}(\partial_u \varphi)^2\right).
\end{eqnarray}
We can also obtain a similar equation for the right-handed fermion.
By taking  derivatives of them, we obtain
\begin{align} 
\nabla^u T_{uu}=& F_{vu} g^{vu}J_u- \frac{c_R \hbar}{48\pi} \partial_u R 
 \nonumber \\
 =& F_{vu} g^{vu}J_u
 - \frac{ \hbar}{48\pi}\left( \frac{c_R-c_L}{2} 
 +\frac{c_L+c_R}{2} \right)  \partial_u R 
 \label{chiral EM left} \\ 
 \nabla^v T_{vv}=&  F_{uv} g^{uv}J_v 
 -\frac{ \hbar}{48\pi}\left( -\frac{c_R-c_L}{2} 
 +\frac{c_L+c_R}{2} \right) \partial_v R
\label{chiral EM right}
\end{align} 
In the case of the non-chiral theory ($c_L = c_R$), we can regard the 
anomalous terms as the contribution of the trace anomaly.  However, in the
case $c_L\ne c_R$, the terms proportional to $(c_L-c_R)$ cannot be regarded
as the contribution of the trace anomaly.  As a result, we obtain the
following anomalous conservation equation and trace anomaly equation:
\begin{align} 
\nabla^\mu T_{\mu\nu}=& 
 F_{\mu\nu} J^\mu 
 - \frac{\hbar}{48\pi} \frac{c_R-c_L}{2} \epsilon_{\mu \nu} \nabla^\mu  R,  \\
{T^\mu}_{\mu}=&\frac{\hbar}{24\pi} \frac{c_L+c_R}{2}R.  
\end{align} 
The first equation reproduces the gravitational anomaly for  the 
covariant EM tensor. 


\subsection{Spin 3 and 4 gauge anomalies}
We have shown that our method reproduces the correct anomaly equations for
 the rank 1 and 2 currents in the chiral theory.
We further consider a generalization to  higher-spin currents.

First we study the rank 3 current.
The equation (\ref{divergence spin 3}) now  becomes
\begin{align} 
 \nabla_v J^{(3)}_{uuu}
 &=-2 F_{uv} J^{(2)}_{uu}
 -\frac{1}{8}\nabla_u\left( g_{uv}R  J^{(1)}_u \right) 
 +\frac{\hbar}{24\pi} 
 \left( \frac{c_R-c_L}{2} +\frac{c_L+c_R}{2} \right) \nabla_u^2 F_{uv}.
 \label{Chiral divergent rank 3}
 \end{align} 
We also obtain a similar equation for $J^{(3)}_{vvv}$.  As in the case of
the energy-momentum tensor, we cannot regard the anomalous term proportional
to $(c_R-c_L)$ as the contribution of the trace anomaly.

Equations consistent with (\ref{Chiral divergent rank 3}) can be given as
follows:
\begin{align}
 \nabla_\mu {J^{(3)\mu}}_{\nu\rho}
 =&-  F_{\nu\mu} {J^{(2)\mu}}_\rho-  F_{\rho\mu} {J^{(2)\mu}}_{\nu}
 -\frac{1}{16}\nabla_\nu\left(R  J^{(1)}_\rho \right)  
 -\frac{1}{16}\nabla_\rho\left(R  J^{(1)}_\nu \right)
 + \frac{1}{16}g_{\nu\rho}\nabla_\mu \left(R J^{(1)\mu} \right) \nonumber \\
 &+ \frac{\hbar}{48\pi}\frac{c_R-c_L}{2}
 \left( 
 \epsilon_{\nu \sigma} \nabla^\sigma \nabla_\mu {F^\mu}_{\rho} 
 + \epsilon_{\rho \sigma} \nabla^\sigma \nabla_\mu {F^\mu}_{\nu}
 - g_{\nu\rho} \epsilon_{\alpha \sigma} \nabla^\sigma 
 \nabla_\mu F^{\mu\alpha}
 \right),
 \\
{J^{(3)\mu}}_{\mu \nu} =& 
 \frac{\hbar}{12\pi}\frac{c_L+c_R}{2}  \nabla_\mu {F^\mu}_{\nu}.
\end{align} 
This is the spin 3 generalization of the gauge or gravitational anomaly.
Note that, the conservation equation has been modified, but the
transformation properties of the background gauge fields
(\ref{transformation rank 3}) - (\ref{transformation rank 1}) are not
changed, since they are classical properties.

We can similarly obtain a generalization to the rank 4 current;
\begin{align}
 \nabla^\mu J^{(4)}_{\mu\nu\rho\sigma} =&  
  F_{\mu\nu} {J^{(3)\mu}}_{\rho \sigma}
  + F_{\mu\rho} {J^{(3)\mu}}_{ \sigma\nu}
  + F_{\mu\sigma} {J^{(3)\mu}}_{\nu\rho } 
  -\frac{1}{8}R
  \left(  
   \nabla_\nu J^{(2)}_{\rho\sigma} 
   + \nabla_\rho J^{(2)}_{\sigma\nu}
   + \nabla_\sigma J^{(2)}_{\nu\rho} 
  \right) 
  \nonumber \\
 & -\frac{1}{6}
  \left( 
   J^{(2)}_{\nu\rho} \nabla_\sigma R 
   +J^{(2)}_{\rho\sigma} \nabla_\nu R
   + J^{(2)}_{\sigma\nu} \nabla_\rho R 
  \right) 
  \nonumber \\
 & -\frac{1}{24} 
  \Bigl( 
  J^{(1)}_{\nu}\nabla_\rho \nabla_\mu {F^\mu}_{\sigma} 
  +J^{(1)}_{\rho}\nabla_\sigma \nabla_\mu {F^\mu}_{\nu}
  +J^{(1)}_{\sigma}\nabla_\nu \nabla_\mu {F^\mu}_{\rho} 
  \nonumber \\
 & +J^{(1)}_{\rho}\nabla_\nu \nabla_\mu {F^\mu}_{\sigma}
  +J^{(1)}_{\nu}\nabla_\sigma \nabla_\mu {F^\mu}_{\rho}
  +J^{(1)}_{\sigma}\nabla_\rho \nabla_\mu {F^\mu}_{\nu}  
  \Bigr) 
  \nonumber \\
& -\frac{\hbar}{960\pi}\frac{c_R-c_L}{2}
\Bigl(
\epsilon_{\nu\alpha} \nabla^\alpha \nabla_\rho \nabla_\sigma R+
\epsilon_{\rho\alpha} \nabla^\alpha \nabla_\sigma \nabla_\nu R+
\epsilon_{\sigma\alpha} \nabla^\alpha \nabla_\nu \nabla_\rho R
 \Bigr)  \nonumber \\
 & -\frac{1}{4}
  \left(
   g_{\nu\rho}\hat{C}_{\sigma}+g_{\rho\sigma}\hat{C}_{\nu}
   +g_{\sigma\nu}\hat{C}_{\rho}  \right),
   \label{spin4gaugeanomaly}
\end{align} 
 \begin{align} 
 {J^{(4)\mu }}_{\mu \nu\rho}
  =-\frac{\hbar}{160\pi}\frac{c_L+c_R}{2} \nabla_\nu \nabla_\rho R
  +g_{\nu\rho}\frac{c_L+c_R}{2}\left[
  \frac{\hbar}{160\pi}\nabla^2 R+\frac{\hbar}{24\pi}\left({\tilde{F}}^2
  -\frac{13}{120}R^2  \right)  \right] .
 \end{align} 
Here we have modified $\tilde{C}_\nu$ to $\hat{C}_\nu$ including the
anomalous terms as follows,
\begin{align} 
 \hat{C}_{\nu} \equiv&  -\frac{1}{4}R \nabla_\rho {J^{(2)\rho}}_{\nu} 
 -\frac{1}{3} {J^{(2)\rho}}_{\nu} \nabla_\rho R 
 -\frac{1}{12} \Biggl(  J^{(1)\rho}\nabla_\rho \nabla_\mu {F^\mu}_{\nu}
 +J^{(1)}_{\rho}\nabla_\nu \nabla_\mu F^{\mu\rho}  \Biggr) \nonumber \\
 &-\frac{\hbar}{960\pi}\frac{c_R-c_L}{2}
\left(   \epsilon_{\nu\alpha} \nabla^\alpha \nabla_\rho \nabla^\rho R+
2\epsilon_{\rho\alpha} \nabla^\alpha \nabla^\rho \nabla_\nu R     \right) .
\end{align} 
This is the spin 4 generalization of the gauge and gravitational
anomalies. The r.h.s. of (\ref{spin4gaugeanomaly}) contains
both of classical and quantum parts.
The classical parts arises due to the same reason as in the 
non-chiral case in section 4. 
The quantum parts are the anomalies.

\section{Summary \label{conclusion}}
\setcounter{equation}{0}

In this paper, we  considered a two-dimensional theory of fermions 
in the electric and gravitational backgrounds and 
obtained a generalization of the gauge, gravitational 
and trace anomalies for higher-spin (HS) currents up to 
spin 4. 
In order to derive these anomalies, we started from
the relation between holomorphic and covariant
forms of HS currents in the electric and gravitational 
backgrounds. 

These anomaly equations can be applied to derive
the higher-spin fluxes of Hawking radiation.
This will be discussed in a separate paper \cite{IMU5}.

In the cases of spins 1 and 2, the forms of anomalies
can be determined by the descent equations and
they have nice geometrical meanings.
It will be interesting to investigate higher-spin
anomalies than 4, and examine whether there are
any systematic structures in the form of anomalies.

Finally we notice that the anomalies we obtained are 
specific to HS currents constructed from fermions.
If they are constructed from bosons, their anomalies have different
combinations with different coefficients.

\appendix

\section{Holomorphic and covariant currents up to spin 4}
\setcounter{equation}{0} 
\label{App-current} 

In section \ref{sec 3}, we represented the covariant currents in terms of
the holomorphic currents as in (\ref{rank3curr}).  However, when we
calculate the conservation equation, it is more convenient to describe the
holomorphic currents in terms of the covariant currents and thus we give
their explicit expressions by expanding the generating function
(\ref{Ghol-cov}). 

\begin{description}
 \item[spin 1 current]
 \begin{align}
  j^{(1)}(u)=:\Psi^\dagger \Psi:=J_u^{(1)}-\frac{1}{\pi} A_u
 \end{align}
 \item[spin 2 current]
 \begin{align}
  j^{(2)}(u)
  =\frac{i}{2}:\Psi^\dagger \partial_u \Psi - \partial_u\Psi^\dagger \Psi:
  = J_{uu}^{(2)} -2A_u J_u^{(1)} +\frac{1}{\pi} A_u^2 
  - \frac{1}{24\pi}
  \left( 
  \partial_u^2\varphi-\frac{1}{2}\left( \partial_u \varphi \right)^2 
  \right)
 \end{align}
 \item[spin 3 current]
\begin{align}
 j^{(3)}(u)=&
 -\frac{1}{4}:\Psi^\dagger \partial_u^2 \Psi 
 - 2\partial_u\Psi^\dagger \partial_u\Psi+\partial_u^2 \Psi^\dagger \Psi: 
 \nonumber \\
 =&J_{uuu}^{(3)} 
 -4A_u J_{uu}^{(2)}
 -\frac{1}{4} \partial_u \varphi \partial_u J_u^{(1)} 
 -\left(
 -4A_u^2+\frac{1}{4}\left(\partial_u^2 \varphi
 -\left( \partial_u \varphi \right)^2 \right)  
 \right) J_u^{(1)} \nonumber \\
 & +\frac{1}{6\pi}
 \left( \partial_u^2 \varphi -\frac{1}{2}(\partial_u \varphi)^2\right) A_u 
 +\frac{1}{12\pi}\partial_u^2 A_u -\frac{4}{3\pi}A_u^3
\end{align}
 \item[spin 4 current]
\begin{align}
j^{(4)}(u) =& \frac{i}{8}:\partial_u^3 \Psi^\dagger \Psi 
 -3 \partial_u^2 \Psi^\dagger \partial_u \Psi
 +3 \partial_u \Psi^\dagger \partial_u^2 \Psi
 - \Psi^\dagger \partial_u^3 \Psi: \nonumber \\
  =&  J^{(4)}_{uuuu}  
  -6A_u J^{(3)}_{uuu} -\frac{3}{4}\partial_u \varphi \partial_u J^{(2)}_{uu}
  - \left[
     \partial_u^2 \varphi -\frac{5}{2}  (\partial_u \varphi)^2
     -12 A_u^2
    \right] J^{(2)}_{uu}
  \nonumber \\
 & +\frac{3}{2} A_u \partial_u \varphi \partial_u J^{(1)}_{u}
  - \left[
    - \frac{3}{2} A_u \left(\partial_u^2 \varphi 
	    -  (\partial_u \varphi)^2\right)
     -\frac{1}{2} \partial_u^2 A_u 
     + 8 A_u^3
    \right] J^{(1)}_{u}
  \nonumber \\
 & -\frac{1}{2\pi}A_u 
\left[ \partial_u^2 A_u 
 + \left(\partial_u^2 \varphi-\frac{1}{2}(\partial_u \varphi)^2 \right) A_u 
 -4 A_u^3 \right]  \nonumber \\ 
 & +\frac{1}{160\pi}\left( \partial_u^4 \varphi 
 - \partial_u \varphi \partial_u^3 \varphi 
 + \frac{4}{3}(\partial_u^2 \varphi)^2 
 -\frac{7}{3}( \partial_u \varphi)^2 \partial_u^2 \varphi
 + \frac{7}{12}( \partial_u \varphi)^4 \right).
\end{align}
\end{description}


\end{document}